\documentclass{article}
\usepackage{graphicx,color}

\newcommand{\be}{\begin{equation}}
\newcommand{\ee}{\end{equation}}
\newcommand{\bdm}{\begin{displaymath}}
\newcommand{\edm}{\end{displaymath}}
\newcommand{\bea}{\begin{eqnarray}}
\newcommand{\eea}{\end{eqnarray}}

\newcommand{\no}{\nonumber \\}

\newcommand{\QCD}{\mbox{\tiny Q\hspace{-0.1em}CD}}

\newcommand{\bc}{\begin{center}}
\newcommand{\ec}{\end{center}}

\newcommand{\ubar}{\overline{\rule[0.42em]{0.4em}{0em}}\hspace{-0.5em}u}
\newcommand{\dbar}{\,\overline{\rule[0.65em]{0.4em}{0em}}\hspace{-0.6em}d}

\newcommand{\sbar}{\overline{\rule[0.42em]{0.4em}{0em}}\hspace{-0.5em}s}

\newcommand{\lvac}{\langle 0|\,}
\newcommand{\rvac}{\,|0\rangle}

\renewcommand{\hbar}{\rule[0.52em]{0.4em}{0.06em}\hspace{-0.45em}h}

\newcommand{\Lo}{\stackrel{\rule[-0.07cm]{0cm}{0cm}\mbox{\tiny LO}}{=}}
\newcommand{\NLo}{\stackrel{\rule[-0.07cm]{0cm}{0cm}\mbox{\tiny NLO}}{=}} 
\newcommand{\lo}{=\hspace{-1.3em}\rule{0em}{0.8em}^{\mbox{\tiny LO}}\hspace{0.2em}}

\begin{document}
\begin{center}
{\large \bf THE MASS OF THE TWO LIGHTEST QUARKS}\footnote{Contribution to the proceedings of the workshop "Determination of the Fundamental Parameters of QCD", Nanyang Technological University, Singapore, 18-22 March 2013, to be published in Mod. Phys. Lett. A.}  \\

\vspace{0.8cm}
H.~Leutwyler\\ Albert Einstein Center for Fundamental Physics\\Institute for Theoretical Physics, University of Bern\\
Sidlerstr.~5, CH-3012 Bern, Switzerland
\end{center}
\vspace{0.2cm}

\begin{abstract}
The problems encountered in the determination of $m_u$ and $m_d$ are discussed. While their sum is known quite well, the difference $m_d-m_u$, which measures the breaking of isospin symmetry in the QCD Lagrangian, is still subject to significant uncertainties. I focus on recent work based on the dispersive analysis of the decay $\eta\rightarrow 3\pi$, which offers a good handle on isospin breaking, because in that transition, the contributions from the e.m. interaction are suppressed. 
\end{abstract}

\section{Standard Model at low energies}	
At low energies ($E \ll\! M_{\mbox{\tiny W}}$), the weak interaction is frozen -- it
only generates tiny effects, visible in the finite lifetime of the particles, for instance. 
Accordingly, the Standard Model reduces to QCD + QED. The parameters in the Lagrangian of this theory are:
 $g,\theta,e$, $m_u$, $m_d$, $m_s$, $m_c$, $m_b$, $m_t$, $m_e$, $m_\mu$, $m_\tau$.
For all we know, this framework provides a precision theory for cold matter ($T\ll\! M_{\mbox{\tiny W}}$). In principle, it allows us to understand the occurrence of nucleons, their mass, their size, their structure, the existence of stable nuclei and their properties, atoms, solids \ldots The Bohr radius, for instance, can be expressed in terms of the above set of parameters: $a_{Bohr}=4\pi/ e^2\,m_e$.

The pattern of quark and lepton masses is not understood at all and looks bizarre indeed. The present talk concerns $m_u$ and $m_d$ -- the least well determined parameters in the above list.  

 \section{Symmetries, effective Lagrangian}  It so happens that $m_u,m_d,m_s$ are small. If these masses would vanish, the Hamiltonian of QCD would be exactly symmetric under the group SU$_{\mbox{\tiny L}}(3)\times$SU$_{\mbox{\tiny R}}(3)$ of chiral transformations: independent flavour rotations of the right- and left-handed components of the three lightest quark  fields. The ground state of the theory, however, is invariant only under the subgroup SU$_{\mbox{\tiny L+R}}(3)$ generated by the vector charges: chiral symmetry is spontaneously broken. As pointed out by Nambu \cite{Nambu,NJL}, the phenomenon generates massless particles, nowadays referred to as {\it Nambu-Goldstone bosons}: if the masses of the three lightest quarks as well as the e.m. coupling constant are sent to zero, the eight lightest mesons, $\pi^-,\pi^0,\pi^+,K^-,K^0,\bar{K}^0,K^+,\eta$ become massless, while the lowest baryons form a degenerate octet of nonzero mass. 
 
 Chiral symmetry is not exact, however. The Hamiltonian of QCD can be decomposed into two parts, $H_{\QCD}= H_0+H_1$,  with $H_1 =\int\! d^3x\, \{m_u\,\ubar u+ m_d \,\dbar d+m_s\,\sbar s\}$. While $H_0$ is invariant under SU$_{\mbox{\tiny L}}(3)\times$SU$_{\mbox{\tiny R}}(3)$, the quark mass term $H_1$ explicitly breaks the symmetry, because it connects the right- and left-handed components. Since $m_u,m_d,m_s$ happen to be small, the matrix elements of $H_1$ are small -- the term can be treated as a perturbation (chiral perturbation theory, $\chi$PT). At leading order of the chiral perturbation series, the pseudoscalar octet is massless. At first order, the square of the pion mass is given by the pion matrix element of the perturbation. As shown by Gell-Mann, Oakes and Renner \cite{GMOR}, chiral symmetry relates this matrix element to the quark condensate:
\be M_{\pi^+}^2\Lo(m_u+m_d)\times|\lvac \ubar u\rvac|\times \frac{1}{F_\pi^{2}}\,.\ee 
 The first term on the right measures the strength of the symmetry breaking in the Hamiltonian, the second is analogous to the spontaneous magnetization of a magnet and represents an order parameter of the spontaneously broken symmetry, while the third is determined by the matrix element $\lvac\dbar \gamma^\mu\gamma_5 u|\pi^+\rangle=i\sqrt{2}\,p^\mu F_\pi$, whose magnitude is known from the pion life time. The symbol LO on top of the equality sign indicates that, as it stands, the relation only holds at leading order -- it receives corrections from higher orders of the chiral perturbation series. 
 
The Gell-Mann-Oakes-Renner formula (1) and its extensions to the other Nambu-Goldstone bosons explain the strange mass pattern at the low energy end of the spectrum: (a) the energy gap of QCD, $M_\pi$, is so small because $m_u,m_d$ happen to be very small, (b) the kaons are much heavier than the pions because it so happens that $m_s\gg m_u,m_d$: the masses of the Nambu-Goldstone bosons very strongly break SU(3)$_{\mbox{\tiny L+R}}$ symmetry because the quark masses do, (c) in contrast to the masses of the NGBs, the matrix elements of the operators $\ubar u$, $\dbar d$, $\sbar s$ do approximately obey the symmetry relations that follow from SU(3)$_{\mbox{\tiny L+R}}$, so that the Gell-Mann-Okubo formula,  $M_\eta^2-M_\pi^2=4(M_\eta^2-M_K^2)$, is obeyed remarkably well.
\section{Quark mass ratios}
The quark mass pattern very strongly breaks isospin symmetry: $m_d$ is about twice as large as $m_u$. The mass splitting between the charged and neutral pions is much smaller. The extension of equation (1) to the $\pi^0$ explains why this is so: it shows that the vacuum shields the pions almost completely from the breaking of isospin symmetry due to the quark mass difference $m_d-m_u$. The mass splitting in the pion multiplet generated by the quark mass difference is proportional to $(m_d-m_u)^2$ and hence tiny -- the observed mass difference stems almost exclusively from electromagnetism.  For this reason, the mean mass, $m_{ud}=\frac{1}{2}(m_u+m_d)$, is more easy to determine than the difference, $m_d-m_u$. 

As pointed out by Weinberg \cite{Weinberg 1977}, the e.m. self energies can be accounted for with Dashen's theorem \cite{Dashen}, which states that the e.m. contributions to the charged Nambu-Goldstone bosons are the same, while those of the neutral particles vanish:
\be M_{K^+}^2\rule[-0.5em]{0.03em}{1.2em}_{\,e.m.}\hspace{-0.3em}\Lo M_{\pi^+}^2\rule[-0.5em]{0.03em}{1.2em}_{\,e.m.}\,, \hspace{2em}M_{K^0}^2\rule[-0.5em]{0.03em}{1.2em}_{\,e.m.}\hspace{-0.3em}\Lo M_{\pi^0}^2\rule[-0.5em]{0.03em}{1.2em}_{\,e.m.}\hspace{-0.3em}\Lo 0\,.\ee
The mass formulae for $M_{\pi^+}, M_{K^0}$ and $M_{K^+}$ can then be solved for the quark mass ratios \cite{Weinberg 1977}
\bea \frac{m_s\rule[-0.1cm]{0cm}{0cm}}{m_{ud}}&\Lo& \frac{M_{K^+}^2 + M_{K^0}^2 - M_{\pi^+}^2}{M_{\pi^0}^2}=25.9\,,\\
\frac{m_u\rule[-0.1cm]{0cm}{0cm}}{m_d}&\Lo&  \frac{M_{K^+}^2 - M_{K^0}^2 + 2 M_{\pi^0}^2 - M_{\pi^+}^2}{M_{K^0}^2 - M_{K^+}^2 + M_{\pi^+}^2} =0.56\,.\eea

As indicated, these low energy theorems are valid only to leading order of the chiral expansion. 
The best estimates for the higher order effects available today are obtained from numerical simulations on a lattice. The estimate for $m_s/m_{ud}$ quoted in the {\it FLAG Review of lattice results concerning low energy particle physics} \cite{FLAG} reads
\be\frac{m_s }{m_{ud}}=27.4\pm 0.4\,,\ee
indicating that this quark mass ratio is now known to an accuracy of 1.5\%. The result shows that the leading term of the chiral perturbation series dominates: the corrections increase the LO term in equation (3) merely by $5.8\pm 1.5\%$.
\section{Low energy theorem for isospin breaking in the meson masses}
The lattice result for $m_s/m_{ud}$ determines the size of the correction in the relation
\be
\frac{M_K^2}{M_\pi^2}= \frac{m_s+m_{ud}}{m_u+m_d}\left\{ 1+\Delta_M\!\right\}\,.\ee
The numerical result (5) implies $\Delta_M=-0.053\,\pm\, 0.013$, indicating that the correction is small also in this case. Remarkably, chiral symmetry implies that the correction of NLO in the ratio of mass splittings is the same \cite{GL 1985}:
\be \frac{M^2_{K^0}-M^2_{K^+}}{M_K^2-M_\pi^2}= \frac{m_d-m_u}{m_s-m_{ud}}\left\{ 1+\Delta_M+
O(m_{q}^2)\right\}\,.
\ee
Hence the quark mass ratio
\be Q^2\equiv\frac{m_s^2-m_{ud}^2}{m_d^2-m_u^2}\ee
is given by a ratio of meson masses, up to corrections of NNLO:
\be Q^2\NLo\frac{M_K^2-M_\pi^2 }{M^2_{K^0}-M^2_{K^+}}\cdot\frac{M_K^2}{M_\pi^2} \,. \ee
Using the Dashen theorem to account for the e.m. corrections, this gives $Q$ = 24.3. 

The quantity $Q$ compares the isospin symmetry breaking parameter $m_d-m_u$ with the quantity $m_s-m_{ud}$, which measures the strength of SU(3)$_{\mbox{\tiny L+R}}$ symmetry breaking. Since the ratio $m_s/m_{ud}$ is accurately known, the value of $Q$ determines the value of $m_u/m_d$ and vice versa.
Most of the results underlying the estimate (5) are obtained from lattice simulations of QCD and hence disregard electromagnetic effects. In the case of $m_s/m_{ud}$, this is a good approximation, because the uncertainties in the corrections to the Dashen theorem barely affect this ratio. For $m_u/m_d$, however, the e.m. interaction does play a significant role. Lattice simulations of QCD + QED cannot be done with the same level of confidence as for QCD alone: for the time being, the e.m. self energies are evaluated in the quenched approximation and the role of the finite size effects in the presence of long range forces yet needs to be studied.  The value of $Q$ quoted in the FLAG review \cite{FLAG}, $Q=22.8\pm 1.2$, relies on estimates of the e.m. corrections derived from $\eta$ decay \cite{GL eta,Kambor Wiesendanger Wyler,Anisovich Leutwyler,Leutwyler CD 2009}. 

In Bern, we have pursued the determination of $Q$ from $\eta$ decay over many years \cite{Walker,Colangelo Lanz Passemar,Lanz PhD,Colangelo Lanz Leutwyler Passemar,Lanz CD 2012}. In the following, I outline recent work done in collaboration with Gilberto Colangelo, Stefan Lanz and Emilie Passemar. A detailed report on this project is forthcoming \cite{CLLP to be published}. 
\section{ $\eta$ decay} 
The decay $\eta\rightarrow 3\pi$ provides a better handle on $Q$ than the mass splitting between $K^+$ and $K^0$, because the e.m.\,interaction is suppressed (Sutherland's theorem \cite{Sutherland,Bell Sutherland}).
In the limit $e=0$, $m_u=m_d$, isospin and hence $G$-parity are conserved. In view of $G_\eta=1,\,G_\pi = -1 $, the transition is then forbidden: the $\eta$ becomes a stable particle. Accordingly, $\eta\rightarrow 3\pi$  is sensitive to isospin breaking and, since the e.m.\,contributions are tiny \cite{Baur Kambor Wyler}, the transition amplitude is to a very good approximation proportional to $m_d-m_u$. In the following, I disregard the e.m. interaction, but will return to it in section 7. 

The structure of the leading term of the chiral perturbation series \cite{Cronin,Osborn Wallace},
\be A(\eta\rightarrow\pi^+\pi^-\pi^0)\Lo -\frac{\sqrt{3}}{4 }\cdot\frac{m_d-m_u}{m_s-m_{ud}}\cdot\frac{s-\frac{4}{3}M_\pi^2}{F_\pi^2}\,\ee
resembles the leading term in the chiral expansion of the $\pi\pi$ scattering amplitude:
\be A(\pi\pi\rightarrow\pi\pi)\Lo\frac{s-M_\pi^2}{F_\pi^2}\,.\ee
 In both cases, the amplitude is linear in $s$ and contains an
Adler zero. In the case of $\pi\pi$ scattering, the zero occurs at 
$s_A\lo M_\pi^2$, while in $\eta$ decay, $s_A\lo \frac{4}{3}M_\pi^2$.
The analytic structure of the two amplitudes is also very similar.
In either case, the higher order contributions of the chiral perturbation series are dominated by the final state interaction among the pions.

The correction of next-to-leading order was worked out long ago \cite{GL eta}, by evaluating the chiral perturbation series to one loop. The most remarkable property of the result is that, expressed in terms of the quark mass ratio $Q$, 
\be  A(\eta\rightarrow\pi^+\pi^-\pi^0)\NLo -\frac{1}{Q^2}\cdot\frac{M_K^2(M_K^2-M_\pi^2)}{3\sqrt{3} M_\pi^2 F_\pi^2}\cdot M(s,t,u)\,,\ee
all of the low energy constants except one drop out: the factor $M(s,t,u)$ exclusively involves $F_\pi$, $F_K$, $M_\pi$, $M_K$, $M_\eta$ and $L_3$.  Moreover, $L_3$ does not concern the dependence of the amplitude on the quark masses, on which there is only indirect experimental information, but the momentum dependence -- the value of $L_3$ can be determined quite well from $\pi\pi$ scattering. At one loop, the result for the rate is therefore of the form $\Gamma_{\eta\rightarrow\pi^+\pi^-\pi^0} =C/Q^4$, where $C$ is a known constant. Hence $Q$ can be determined from the observed rate. 

The main problem in this determination of $Q$ is not the uncertainty in $L_3$, but concerns the contributions from higher orders. In 1985, we estimated the uncertainty in the result  at 
$Q^{-2}=(1.9\pm0.3)\cdot 10^{-3}$, which amounts to $Q =22.9^{+2.1}_{-1.6}$. 
This is consistent with the value $Q=24.3$ obtained from the kaon mass difference with the Dashen theorem, but the uncertainties are large.

\section{Dispersive analysis of $\eta$ decay} 
The properties of the decay amplitude are governed by the final state interaction among the three pions. Up to and including NNLO of the chiral perturbation series, the amplitude can be represented in terms of three functions of a single variable \cite{Fuchs Sazdjian Stern 1993}:
\be M(s,t,u)=M_0(s)+(s-u)M_1(t)+(s-t)M_1(u)+M_2(t)+M_2(u)-\mbox{$\frac{2}{3}$}M_2(s)\ee
(discontinuities from partial waves with $\ell\geq2$ start contributing only at $\mbox{N}^3$LO). Unitarity implies that $M_0(s)$, $M_1(s)$, $M_2(s)$ have a branch cut extending from $4M_\pi^2$ to $\infty$. In the elastic region, the discontinuity across the cut is determined by the S- and P-wave phase shifts of $\pi\pi$ scattering. Neglecting the discontinuities due to inelastic processes, the dispersion relations obeyed by the three functions can be brought to the form \cite{Anisovich Leutwyler}
\be M_I(s)=\Omega_I(s)\left\{P_I(s)+\frac{s^{n_I}}{\pi}\int_{4M_\pi^2}^\infty ds'\frac{\sin\delta_I(s')\hat{M}_I(s')}{|\Omega_I(s')|s^{\prime\, n_I}(s'-s)}\right\},\hspace{0.4cm}(I=0,1,2),\ee
 where $\delta_0(s), \delta_1(s),\delta_2(s)$ are the S- and P-wave phase shifts of $\pi\pi$ scattering,
\be \Omega_I(s)\equiv\exp\left\{\frac{s}{\pi}\int_{4M_\pi^2}^\infty ds'\frac{\delta_I(s')}{s'(s'-s)}\right\},\hspace{0.4cm}(I=0,1,2)\ee
is the corresponding Omn\`{e}s factor and the polynomials $P_0(s), P_1(s),P_2(s)$ collect the subtraction constants. The function $\hat{M}_I(s)$ denotes an angular average -- it arises from scattering in the crossed channels, $s\leftrightarrow t$, $s\leftrightarrow u$. Formally, the dispersion integrals extend to $\infty$, but with the number of subtractions we are using, the contributions from the discontinuities above $K\bar{K}$ threshold are too small to matter.

The situation is quite similar to the one for $\pi\pi$ scattering. The main difference is that the subtraction constants relevant for $\eta\rightarrow 3\pi$ cannot be predicted to the same precision. 
While $\pi\pi$ scattering can be analyzed within the effective theory built on SU(2)$_{\mbox{\tiny L}}\times$SU(2)$_{\mbox{\tiny R}}$, which treats only $m_u$ and $m_d$ as small, the theoretical estimates of the subtraction constants relevant for $\eta$-decay rely on SU(3)$_{\mbox{\tiny L}}\times$SU(3)$_{\mbox{\tiny R}}$ and hence treat $m_s$ as an expansion parameter as well. Only the occurrence of an Adler zero follows from SU(2)$_{\mbox{\tiny L}}\times$SU(2)$_{\mbox{\tiny R}}$ symmetry alone. 

The fact that the $\eta$ is not a stable particle implies that the evaluation of the integrands occurring in the dispersion relations (14) is not trivial. A coherent framework is obtained with analytic continuation in the mass of the $\eta$. Explicit expressions for the relevant angular integrals, together with a detailed discussion of the steps required to analytically continue these in $M_\eta$ were given in \cite{Anisovich}. 

The dispersion relations (14) are linear in the decay amplitude: if $M^{(1)}(s,t,u)$ and $M^{(2)}(s,t,u)$ are solutions, then $\lambda_1M^{(1)} +\lambda_2M^{(2)} $ is one as well. Hence the solutions form a linear space \cite{AL unpublished}: the amplitude can be represented as a linear superposition of basis functions (the number of independent solutions depends on the number of subtractions made). Since the basis functions can be calculated once and for all, this property simplifies the comparison with the data considerably. 

As was to be expected, the dispersive treatment amplifies the final state interaction effects occurring in the one loop representation of $\chi$PT, but the modification is quite modest. This indicates that, throughout the region relevant for our analysis, $0\leq s,t,u\leq(M_\eta-M_\pi)^2$, the chiral expansion is under good control. The values $Q=22.4 \pm 0.9$ [KWW \cite{Kambor Wiesendanger Wyler}] and  $Q = 22.7 \pm 0.8$ [AL \cite{Anisovich Leutwyler}] found in 1996 not only confirmed the result  $Q =22.9^{+2.1}_{-1.6}$ [GL \cite{GL eta}] obtained earlier (directly from the one loop representation), but also reduced the uncertainty by a factor of 2.

 \section{Recent work on $\eta$ decay}
A thorough analysis of the ingredients needed in the determination of $Q$ from $\eta$ decay indicated the need for further work  \cite{Bijnens Gasser}, in particular also on the experimental side. In the meantime, the experimental information on $\eta\rightarrow 3\pi$  improved enormously, on account of the work done at KLOE, MAMI and WASA. Andrzej Kupsc (KLOE), Sergey Prakhov (MAMI) and Patrik Adlarson (WASA) kindly provided us with detailed data tables. The uncertainties, not only in the Dalitz plot distributions, but even in the decay rates, which posed a serious limitation in early work, have practically disappeared. In particular, the compilation provided by the Particle Data Group \cite{PDG} shows that the experimental information about the slope of the neutral Dalitz plot is now in very good shape. 

At the precision reached, isospin breaking needs to be accounted for. In particular, the presence of charged particles in the final state requires radiative corrections. Moreover, the e.m. self energy of the pions generates a sizeable difference between the masses of the charged and neutral pions, which affects the phase space integrals quite significantly. A complete calculation in the effective theory of QCD + QED has now been carried out to NLO of the chiral expansion \cite{Ditsche Kubis Meissner}. We rely on this work to account for the e.m. effects. The fact that the value of $Q$ can be determined either from the rate of the transition $\eta\rightarrow \pi^+\pi^-\pi^0$ or from $\eta\rightarrow 3\pi^0$ offers a good test: evaluating the e.m. corrections on the basis of the one loop representation, we find that the two results indeed agree.
 
For $m_u=m_d$ and  $e=0$, the chiral perturbation series of the amplitude is now known to two loops \cite{Bijnens Ghorbani}. The main problem encountered when comparing this representation of the amplitude with experiment is the occurrence of a plethora of  low energy constants, only some of which can reliably be estimated. In particular, it is notoriously difficult to estimate those LECs that control the dependence on the quark masses, because direct experimental information about that is not available. The relevant sum rules receive contributions from scalar intermediate states, which cannot be estimated with resonance saturation -- while vector meson dominance is often an adequate approximation, scalar meson dominance fails. For a recent discussion of some of the problems encountered in the comparison of the two loop representation with data, I refer to a paper by Kolesar \cite{Kolesar}.

A different development concerns the analysis of the decay $\eta\rightarrow 3\pi$ within the nonrelativistic effective theory \cite{Gullstrom Kupsc Rusetsky,Kupsc Rusetsky Gullstrom,Schneider Kubis Ditsche}, analogous to the one successfully used for the analysis of $K\rightarrow 3\pi$ \cite{CGKR,Bissegger 1,Bissegger 2,Bissegger 3}.  This framework, in particular allows one to study the relation between the behaviour of the Dalitz plot distributions of the charged and neutral decay modes in the vicinity of the centre of the plot.

I briefly comment on an entirely different approach, which recently appeared in print \cite{Kampf:2011wr}. The ingredients of that work are very similar to ours: dispersion theory and experimental information are used to improve the representations obtained in the framework of $\chi$PT. The result is very different from ours, however. 
The difference is most clearly seen in the behaviour of the real part of the amplitude along the line $s=u$. In the physical region, the amplitude constructed by these authors is not very different from ours, but below threshold, in the region $0\leq s\leq 4M_\pi^2$, there is a qualitative difference: while our representation stays close to the linear leading order formula in equation (10) and hence passes through an Adler zero in the vicinity of $s_A=\frac{4}{3}M_\pi^2$, theirs bends upwards -- it does not contain an Adler zero at all. 

A low energy theorem of SU(2)$_{\mbox{\tiny L}}\times$SU(2)$_{\mbox{\tiny R}}$ states that, in the limit $m_u=m_d=0$, the amplitude vanishes in two corners of the Mandelstam triangle: $s=u=0$ and $s=t=0$. If the quark masses are turned on, the zeros are pushed inside the triangle, by an amount proportional to $m_u+m_d$. At leading order of the chiral perturbation series, the amplitude vanishes along the line $s_A\lo\frac{4}{3}M_\pi^2$ (see section 5). The contributions of NLO modify the line where the real part of the amplitude vanishes, but the modification is very modest: the value of $s_A$ (value of $s$ at which the function $\mbox{Re}\hspace{0.01cm}M(s,t,u)\hspace{0.02cm}\rule[-0.2em]{0.03em}{1em}_{\,s=u}$ passes through zero) increases by $6.4\%$. The NLO correction to the low energy theorem for the slope at the Adler zero, $ D_A=\partial_s\hspace{-0.1cm}\left\{\mbox{Re}\hspace{0.01cm}M(s,t,u)\hspace{0.02cm}\rule[-0.2em]{0.03em}{1em}_{\,s=u}\right\}$, is of similar size: the leading order prediction, $D_A\lo 3/(M_\eta^2-M_\pi^2)$, is increased by $6.5\%$. The NNLO representation also contains a zero in the immediate vicinity of $\frac{4}{3}M_\pi^2$, etc. It is true of course that chiral symmetry is only an approximate symmetry, but it appears to me that a calculation which invokes results obtained from $\chi$PT and comes up with a representation that is in conflict with one of the key consequences of the fact that the pions are the Nambu-Goldstone bosons generated by the spontaneous breakdown of this symmetry cannot be internally consistent. 

In our work, we assume that the amplitude does contain an Adler zero. Since we do not know why the corrections are significantly smaller than the typical size of SU(3)$_{\mbox{\tiny L}}\times$SU(3)$_{\mbox{\tiny R}}$ breaking effects, we use the standard estimate for the uncertainties to be attached to NLO results, for $s_A$ as well as $D_A$.  

\section{Subtraction constants}
In the form specified in equation (14), the dispersion relations uniquely fix the amplitude in terms of the subtraction constants and the $\pi\pi$ phase shifts \cite{AL unpublished}. The latter are now known to remarkable accuracy, due to a combined effort on the experimental and theoretical sides: low energy precision experiments on $K_{\ell4}$ decays (E865, NA48, DIRAC) have led to an accurate experimental determination of the S-wave scattering lengths, which beautifully confirms the theoretical predictions (see \cite{Balev:2011zz} for a recent review). The scattering lengths play a crucial role because they determine the subtraction constants needed in the dispersive analysis of $\pi\pi$ scattering (Roy equations). In view of these developments, the uncertainties in the phase shifts do not play a significant role any more in the determination of $Q$ from $\eta$ decay: dispersion theory fixes the decay amplitude in terms of the subtraction constants within very narrow limits. 

We allow for altogether eleven subtractions, using cubic polynomials for $P_0(s), P_2(s)$ and a quadratic one for $P_1(s)$. Not all of the subtraction constants are of physical significance, however, because the decomposition of the amplitude in equation (13) is not unique: five of the eleven constants can be modified at will -- if the six remaining ones are properly adjusted, the sum over the isospin components remains the same.  

The data on the Dalitz plot distributions strongly constrain the values of the physically relevant subtraction constants, but cannot fully determine them, because the data do not constrain the magnitude of the amplitude at the centre of the Dalitz plot. Theoretical information obtained within the effective theory is indispensable to determine the normalization. We assume that the one loop representation of $\chi$PT represents a good approximation, not at the centre of the Dalitz plot, but at small values of $s,t,u$, where the higher orders of the chiral perturbation series are smallest. The isospin components of the amplitude are expanded in a Taylor series:
\bea M_0(s)&=& a_0+ b_0 s + c_0 s^2+ d_0 s^3\ldots\no
M_1(s)&=& a_1+ b_1 s +c_1 s^2 \ldots\\
M_2(s)&=& a_2+ b_2 s + c_2 s^2+ d_2 s^3\ldots\nonumber\eea 
Since the Omn\`es factors are complex, the subtraction polynomials in equation (14) need not be real, but the chiral expansion of the Taylor coefficients shows that  these are real, up to and including NLO. An imaginary part starts showing up only at two loops. In fact, the explicit expression does not involve any unknown LECs, so that the imaginary parts of the Taylor coefficients can be evaluated numerically without further ado. The result is different from zero, but very small: while the subtraction constants are complex, the Taylor coefficients are approximately real. It makes very little difference whether we take their imaginary parts from $\chi$PT or set them equal to zero. The one loop representation yields a parameter free estimate for the real parts of $a_0,b_0,c_0,a_1,b_1,a_2,b_2,c_2$. We estimate the uncertainties due to higher order terms in the chiral perturbation series in the standard way. As we are dealing with SU(3)$_{\mbox{\tiny L}}\times$SU(3)$_{\mbox{\tiny R}}$ we use the typical size of SU(3) symmetry breaking effects: 20 to 30\%  at LO and the square of that at NLO.  
 
As a side remark, I mention that the two loop representation in addition also specifies the three remaining coefficients $d_0,c_1,d_2$ in equation (16), but in view of the unknown LECs, the information flows in the opposite direction: we can use the dispersive analysis to estimate some of the low energy constants occurring at two loops.  In our analysis, $d_0,c_1,d_2$ are treated as free parameters, to be determined with the measured Dalitz plot distributions.  

This completes the outline of our analysis. I refrain from quoting preliminary numerical results because the error analysis yet needs to be completed. A detailed account is in preparation \cite{CLLP to be published}.  

\section*{Acknowledgment}
It is a pleasure to thank J.~Gasser, A.V.~Anisovich and M.~Walker for collaboration in an early phase of the analysis described here and G.~Colangelo, S.~Lanz and E.~Passemar for making an update of this work possible. Also, I thank P.~Adlarson, A.~Kupsc and S.~Prakhov for providing us with the data base used in our calculation. Informative discussions with A.~Rusetsky and B.~Kubis, who made the analytic representations of some of their  results accessible to us, were most useful. Finally, I wish to thank Prof.~Phua for kind hospitality at the Institute for Advanced Studies, Nanyang Technological University, Singapore.


\begin{thebibliography}{0}
\bibitem{Nambu} Y.~Nambu,
  %``Axial vector current conservation in weak interactions,''
  {\it Phys. Rev. Lett.}  {\bf 4} (1960) 380.
\bibitem{NJL}
 Y.~Nambu and G.~Jona-Lasinio,
  %``Dynamical Model of Elementary Particles Based on an Analogy with Superconductivity. 1.,''
  {\it Phys. Rev.}  {\bf 122} (1961) 345.

\bibitem{GMOR} M.~Gell-Mann, R.~J.~Oakes and B.~Renner,
  %``Behavior of current divergences under SU(3) x SU(3),''
  {\it Phys. Rev.} {\bf 175} (1968) 2195.

\bibitem{Weinberg 1977}
S.\ Weinberg, in {\it A Festschrift for I.I. Rabi}, ed.\
L.\ Motz, {\it Trans. New York Acad. Sci.} Ser. II {\bf 38} (1977) 185.

\bibitem{Dashen}R. Dashen, {\it Phys. Rev.} 183 (1969) 1245.

\bibitem{FLAG}G.~Colangelo {\it et al.},
  %``Review of lattice results concerning low energy particle physics,''
  {\it Eur.Phys.\ J.}\ C {\bf 71} (2011) 1695 [arXiv:1011.4408].

\bibitem{GL 1985} J.~Gasser and H.~Leutwyler,
  %``Chiral Perturbation Theory: Expansions in the Mass of the Strange Quark,''
  {\it Nucl.\ Phys.}\ B {\bf 250} (1985) 465.
   
 \bibitem{GL eta}J.~Gasser and H.~Leutwyler,
  %``eta ---> 3 pi to One Loop,''
 {\it  Nucl.\ Phys.}\ B {\bf 250} (1985) 539.

\bibitem{Kambor Wiesendanger Wyler}J.~Kambor, C.~Wiesendanger and D.~Wyler,
  %``Final state interactions and Khuri-Treiman equations in eta --> 3 pi decays,''
  {\it Nucl.\ Phys.}\ B {\bf 465} (1996) 215.
  
\bibitem{Anisovich Leutwyler}A.~V.~Anisovich and H.~Leutwyler,
  %``Dispersive analysis of the decay eta ---> 3 pi,''
  {\it Phys.\ Lett.}\ B {\bf 375} (1996) 335.
 
\bibitem{Leutwyler CD 2009}
H.~Leutwyler,
  %``Light quark masses,''
  PoS CD {\bf 09} (2009) 005
  [arXiv:0911.1416].
  
\bibitem{Walker} 
M. Walker, %$\eta\rightarrow 3\pi$, 
diploma thesis, University of Bern, 1998.

\bibitem{Colangelo Lanz Passemar}
G.~Colangelo, S.~Lanz and E.~Passemar,
  %``A New Dispersive Analysis of eta ---> 3 pi,''
{\it  PoS CD\bf 09} (2009) 047
  [arXiv:0910.0765].
 
 \bibitem{Lanz PhD} 
S. Lanz, %Determination of the quark mass ratio $Q$ from $\eta\rightarrow 3\pi$, 
PhD thesis, University of Bern, 2011.

\bibitem{Colangelo Lanz Leutwyler Passemar}
G.~Colangelo {\it et al.},
  %``Determination of the light quark masses from eta ---> 3pi,''
  {\it PoS EPS-HEP2011} (2011) 304. 

\bibitem{Lanz CD 2012}
S.~Lanz,  %``Eta --> 3 pi and quark masses,''
{\it Proceedings Chiral Dynamics 2012},  arXiv:1301.7282.   

\bibitem{CLLP to be published}G.~Colangelo, S.~Lanz, H.~Leutwyler and E.~Passemar, in preparation. 
  
\bibitem{Sutherland}D.~G.~Sutherland,
  %``Current algebra and the decay eta ---> 3pi,''
  {\it Phys.\ Lett.}\  {\bf 23} (1966) 384.
  
\bibitem{Bell Sutherland}  
  J.~S.~Bell and D.~G.~Sutherland,
  %``Current algebra and eta ---> 3 pi,''
  {\it Nucl.\ Phys.}\ B {\bf 4} (1968) 315.
 
\bibitem{Baur Kambor Wyler}R.~Baur, J.~Kambor and D.~Wyler,
  %``Electromagnetic corrections to the decays eta --> 3 pi,''
  {\it Nucl.\ Phys.}\ B {\bf 460} (1996) 127
  [hep-ph/9510396].

 \bibitem{Cronin}J.~A.~Cronin,
  %``Phenomenological model of strong and weak interactions in chiral U(3) x U(3),''
  {\it Phys.\ Rev.}\  {\bf 161} (1967) 1483.

\bibitem{Osborn Wallace}
H.~Osborn and D.~J.~Wallace,
  %``Eta - x mixing, eta ---> 3pi and chiral lagrangians,''
  {\it Nucl.\ Phys.}\ B {\bf 20} (1970) 23.
 
\bibitem{Fuchs Sazdjian Stern 1993}
J.~Stern, H.~Sazdjian and N.~H.~Fuchs,
  %``What pi - pi scattering tells us about chiral perturbation theory,''
{\it   Phys.\ Rev.}\ D {\bf 47} (1993) 3814
  [hep-ph/9301244].

\bibitem{Anisovich}
A.~V.~Anisovich,
  %``Dispersion relation technique for three pion system and the P wave interaction in eta ---> 3 pi decay,''
{\it  Phys.\ Atom.\ Nucl.}\  {\bf 58} (1995) 1383
   [Yad.\ Fiz.\  {\bf 58N8} (1995) 1467].
   
\bibitem{AL unpublished}A.~V.~Anisovich and H.~Leutwyler, unpublished work (1996).

\bibitem{Bijnens Gasser}J.~Bijnens and J.~Gasser,
  %``Eta decays at and beyond p**4 in chiral perturbation theory,''
 {\it  Phys.\ Scripta} T {\bf 99} (2002) 34
  [hep-ph/0202242].
 
\bibitem{PDG}J.~Beringer {\it et al.}  [Particle Data Group Collaboration],
  %``Review of Particle Physics (RPP),''
{\it  Phys.\ Rev.}\ D {\bf 86}, 010001 (2012).

\bibitem{Ditsche Kubis Meissner}C.~Ditsche, B.~Kubis and U.-G.~Mei\ss ner,
  %``Electromagnetic corrections in eta ---> 3 pi decays,''
  {\it Eur.\ Phys.\ J.}\ C {\bf 60} (2009) 83 
  
  [arXiv:0812.0344]. 
  
\bibitem{Bijnens Ghorbani}
J.~Bijnens and K.~Ghorbani,
  %``eta ---> 3 pi at Two Loops In Chiral Perturbation Theory,''
{\it  JHEP} {\bf 0711} (2007) 030
  [arXiv:0709.0230].

\bibitem{Kolesar}M.~Kolesar,
  %``Analysis of discrepancies in Dalitz plot parameters in eta to 3 pion decay,''
{\it  Nucl.\ Phys.\ Proc.\ Suppl.}\  {\bf 219-220} (2011) 292
  [arXiv:1109.0851].
  
\bibitem{Gullstrom Kupsc Rusetsky}
  C.-O.~Gullstrom, A.~Kupsc and A.~Rusetsky,
  %``Predictions for the cusp in eta ---> 3pi0 decay,''
{\it  Phys.\ Rev.}\ C {\bf 79} (2009) 028201

  [arXiv:0812.2371].

\bibitem{Kupsc Rusetsky Gullstrom}
  A.~Kupsc, A.~Rusetsky and C.-O.~Gullstrom,
  %``A step towards systematic studies of the cusp in eta ---> 3pi0 decay,''
{\it   Acta Phys.\ Polon.\ Supp.}\  {\bf 2} (2009) 169.
  
\bibitem{Schneider Kubis Ditsche}
S.~P.~Schneider, B.~Kubis and C.~Ditsche,
  %``Rescattering effects in eta --> 3pi decays,''
{\it  JHEP} {\bf 1102} (2011) 028
  [arXiv:1010.3946].
    
\bibitem{CGKR}G.~Colangelo, J.~Gasser, B.~Kubis and A.~Rusetsky,
  %``Cusps in K ---> 3 pi decays,''
{\it  Phys.\ Lett.}\ B {\bf 638} (2006) 187
  [hep-ph/0604084].    
  
\bibitem{Bissegger 1}M.~Bissegger, A.~Fuhrer, J.~Gasser, B.~Kubis and A.~Rusetsky,
  %``Cusps in K(L) ---> 3 pi decays,''
{\it  Phys.\ Lett.}\ B {\bf 659} (2008) 576
  [arXiv:0710.4456].

\bibitem{Bissegger 2} M.~Bissegger, A.~Fuhrer, J.~Gasser, B.~Kubis and A.~Rusetsky,
  %``Radiative corrections in K ---> 3 pi decays,''
{\it  Nucl.\ Phys.}\ B {\bf 806} (2009) 178
  [arXiv:0807.0515].
  
\bibitem{Bissegger 3}J.~Gasser, B.~Kubis and A.~Rusetsky,
 %``Cusps in K --> 3pi decays: a theoretical framework,''
 Nucl.\ Phys.\ B {\bf 850} (2011) 96
 [arXiv:1103.4273].
   
\bibitem{Kampf:2011wr}
  K.~Kampf, M.~Knecht, J.~Novotny and M.~Zdrahal,
  %``Analytical dispersive construction of $\eta\to3\pi$ amplitude: first order in isospin breaking,''
{\it  Phys.\ Rev.}\ D {\bf 84} (2011) 114015
  [arXiv:1103.0982].  

\bibitem{Balev:2011zz}
  S.~Balev {\it et al.}  [NA48-2 Collaboration],
  %``Test of chiral perturbation theory with K+-(e4) and K00(e4) decays at NA48/2,''
{\it  PoS EPS-HEP2011} (2011) 437.
 
 \end{thebibliography}
\end{document}